\documentclass[twocolumn, 10pt]{article}
\usepackage[cm]{fullpage}

\usepackage{amssymb,amsfonts,times,url}
\usepackage{graphicx}
\usepackage[round]{natbib}
\begin{document}

\title{An Alternative Model of Amino Acid Replacement}

\author{Gavin E. Crooks\footnote{To whom correspondence should be addressed. gec@compbio.berkeley.edu}~\ and Steven E. Brenner \\
Dept. of Plant and Microbial Biology, 111 Koshland Hall \#3102, \\
University of California,\\
Berkeley, CA 94720-3102, USA
}
\maketitle

\begin{abstract}
%
The observed correlations between pairs of homologous protein sequences are typically explained in terms of a Markovian dynamic of amino acid substitution. This model assumes that every location on the protein sequence has the same background distribution of amino acids, an assumption that is incompatible with the observed heterogeneity of protein amino acid profiles and with the success of profile multiple sequence alignment. 
%
We propose an alternative model of amino acid replacement during protein evolution based upon the assumption that  the variation of the amino acid background distribution from one residue to the next is sufficient  to explain the observed sequence correlations of homologs. The resulting dynamical model of independent replacements drawn from heterogeneous backgrounds is simple and consistent, and provides a unified homology match score for sequence-sequence, sequence-profile and profile-profile alignment.
%
\end{abstract}

\section*{Introduction}

During evolution, a protein's amino acid sequence is altered by the insertion and deletion of residues and by the  replacement of one residue by another.
In principle, the alignment of proteins sequences and the subsequent detection of protein homologs and the inference of protein phylogenies requires a dynamical model of this sequence evolution. 
The most common and widely used residue replacement dynamics is the
 standard Dayhoff model, which  assumes that the substitution probability during some time interval depends only on the identities of the initial and replacement residues and that the dynamics is otherwise homogeneous along the protein chain, and between protein families, and across evolutionary epochs. 
 In other words, under this model the dynamics of  amino acid substitution resembles a continuous time, first order Markov chain \citep{Dayhoff1972,Dayhoff1978, Gonnet1992, Jones-1992-CABIOS, Muller-2000-JCompBiol}. 

However, it has long been known that this widely used Markovian substitution model  is fundamentally unsatisfactory.
One major problem is that the short and long time substitution dynamics are incompatible \citep{Gonnet1992,Benner1994, Muller-2000-JCompBiol}. \citet{Benner1994} suggests that this is because  at short evolutionary times the patterns of substitution are influenced by single base mutations between neighboring codons, whereas for more diverged sequences the genetic code is irrelevant and the patterns of replacement are dominated by the selection of  chemically and structural compatible residues.

\begin{figure}[t]
\begin{center}
\includegraphics{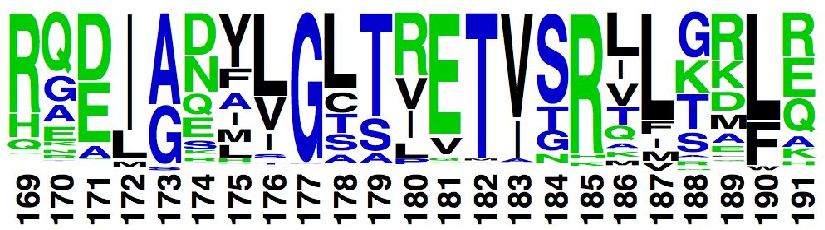}
\caption{The amino acid background distribution of a site within a protein is often stable across large evolutionary time-scales, but varies markedly form one site to another. This figure illustrates the helix-turn-helix motif from the CAP family of homodimeric DNA binding proteins. The height of each letter corresponds to the amino acid frequency in a multiple alignment of 100 diverse, homologous sequences. \citep[For details, see][]{Crooks-2004-GR}
\citep{Schneider-1990-NAR}
These  background distributions are determined by structural, functional and evolutionary constraints.
For example,  positions 180, 181 and 185  are critical to the sequence specific binding of the protein to DNA, the conserved glycine at position 177 is located on inside of the turn between the helices, and the buried sites  172, 176, 178, 183, 187 and 190 contain mostly hydrophobic residues.
It should be noted that the correlations inherent in these background distributions are far stronger than can be explained by local structural features (such as burial and secondary structure) alone. \citep{Crooks2004a,Crooks2004b}.
}
\label{cap}
\end{center}
\end{figure}

A more serious problem with the Dayhoff Markovian model  is that it assumes that every residue in every protein has the same background distribution of amino acids and that protein sequences rapidly evolve to this uninteresting equilibrium.
In actuality, the amino acid background distribution varies markedly from one residue position to the next, as can be seen, by example, in figure~\ref{cap}. These  large site-to-site variations are stable across relatively long evolutionary time-scales, 
and they account for the success of protein hidden Markov models and other profile based multiple sequence alignment methods. \citep[See, for example,][]{Sjolander-1996-CABIOS,DurbinEddy1998} 
Profile methods can detect substantially more remote homologies than pairwise alignment \citep[][Green and Brenner, Unpublished data]{Park-1998-JMB}. 
In short, the dynamics of amino acid substitution are not Markovian, stationary, nor homogeneous, and the prediction of rapidly decaying sequence correlations is at odds with the success of profile based remote homology detection. 

A natural solution to the limitations of the Markov model is to assume that residue replacement is governed by different Markov processes for each position, each process potentially possessing its own background distribution and substitution probabilities. The appropriate Markov matrix for a particular protein position is chosen based upon predictions of the protein structure, or directly from the sequence data. \citep{Goldman1996,Thorne1996,Topham1997,Goldman1998, Koshi1998,Dimmic2000,Lartillot2004}  However, this approach is both computationally and conceptually complex.

\begin{figure}[t]
\begin{center}
\includegraphics{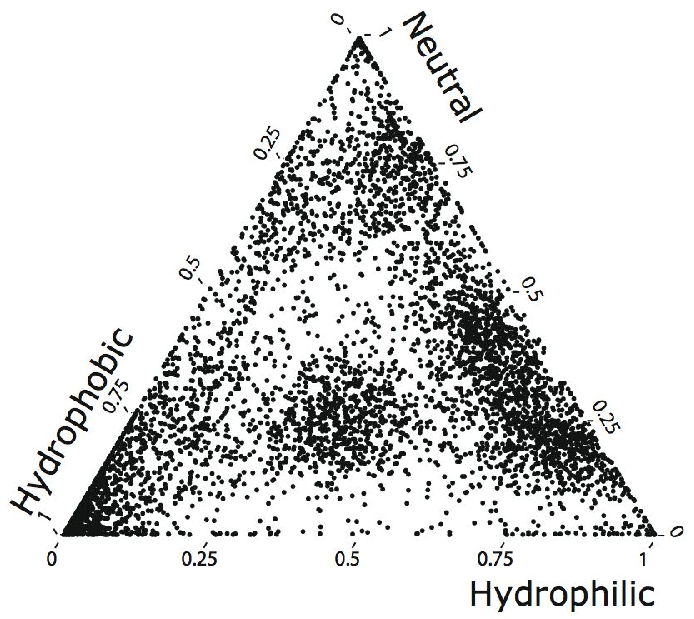}
\caption{ The distribution of amino acid backgrounds is heterogeneous and multimodal. 
This ternary scatter plot represents 5000 randomly sampled distributions
drawn from the {\tt dist.20comp} Dirichlet mixture model  \citep{Karplus-1995-TN-Regularizers} of  amino acid backgrounds. Each has been
 projected onto the three dimensional subspace of hydrophobic (C, F, I, L, M, V, W, Y), neutral (A, G, P, S, T) and hydrophilic (D, E, H, K, N, Q, R) residues. 
Major peaks in the probability density are located around (0.5, 0.25, 0.25)  hydrophobic/neutral/hydrophilic, and towards the vertices of this ternary plot. Thus, if we observe several homologous hydrophilic residues  (e.g. position 179 or 184 in fig.~\ref{cap}) we can be reasonably confident that additional homologous residues will also be hydrophilic.
}
\label{dirichlet}
\end{center}
\end{figure}

Here, we propose that the observed sequence correlation between diverged homologs is principally due to the heterogeneous, stable, background distribution of each protein site and, therefore, that a Markovian amino acid replacement dynamics is overly complicated and possible unnecessary for the accurate construction of  protein sequence alignments and phylogenies. As an alternative, we construct a dynamical model of amino acid replacement that explicitly assumes that each protein site has a different equilibrium distribution of the 20 canonical amino acids (which we refer to as that site's amino acid background, $\theta$) and that the residue distribution of each site rapidly (relative to evolutionary time scales) relaxes to this local, site-specific equilibrium.
These distributions themselves conform to a probability distribution of backgrounds, $P(\theta)$,  which we may discover by studying many families of homologous proteins (See Fig.~\ref{dirichlet}). 
We do not need to model the short time dynamics
with any great accuracy, since the alignment of highly conserved homologs is relatively straightforward.
Therefore, we will assume that replacement residues are randomly sampled from the 
background distribution of that site.
Consequentially, (and in direct contrast to the Markov model) the replacement residue is conditionally independent of the initial amino acid at all times.
Note that multiple substitutions are statistically equivalent to a single substitution (since a single mutation is sufficient to relax a site to local equilibrium) and that a residue can be replaced by the same amino acid type.
The resulting dynamic is a site specific, continuous time, zeroth order Markov chain, similar in spirit to Felsenstein's (1981) model of nucleic acid substitution.\nocite{Felsenstein1981} 
The crucial difference is that the initial and replacement residues are non-trivially correlated because both have been sampled from the same background distribution, whereas non-homologous residues are drawn from different backgrounds. 
Bruno has used essentially the same dynamical model discussed here, albeit without incorporating the background prior distribution, to find maximum likelihood estimates of site-specific amino acid frequencies \citep{Bruno1996}.

Under our residue replacement model, the principle  origin of  sequence correlation between diverged homologs is the background distribution of each protein site.  This is also the central idea underlying profile based multiple sequence alignment algorithms. 
Therefore, we are not proposing a radically different method for homolog detection or sequence alignment; rather we are proposing a concrete and consistent dynamics for the underlying evolutionary process.
The implications of this dynamics can be readily extended to cover not only profile-sequence based alignment, but also profile-profile and pairwise sequence-sequence alignment.
Moreover, when we consider pairwise, sequence alignment below, we find that our model is essentially equivalent to the standard pairwise alignment methods, as they are used in practice.
This  alternative dynamical model of amino acid replacement is biologically reasonable, conceptually straightforward and can adequately explain many of the observed patterns of homolog sequence correlation without invoking a Markovian dynamic.

\begin{figure}[tb!]
\begin{center}
\includegraphics{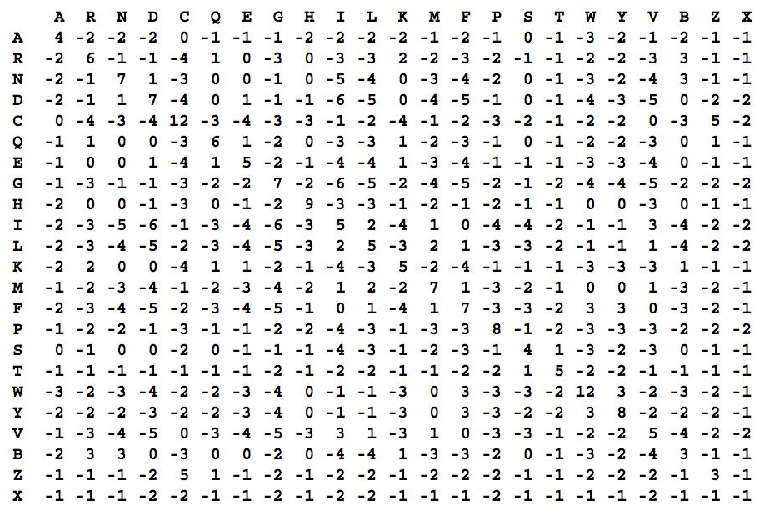}
\caption{
 This substitution matrix (Eq.~\ref{submatrix}, a conventional description of amino acid replacement propensities)
has been directly constructed from the {\tt dist.20comp} Dirichlet mixture model of amino acid background probabilities \citep[Fig.~2;][]{Karplus-1995-TN-Regularizers}, using the conditionally independent substitution model of  amino acid replacement (Eqs.~3-10). 
As a consequence of the heterogeneity and stability of amino acid background probabilities -- illustrated in figs.~\ref{cap} and~\ref{dirichlet} --   the amino acid identity of a pair of alignable, homologous residues is non-trivially correlated over long evolutionary time scales, simply because both residues are drawn from the same background.
Scores are in units of $\frac{1}{3}$ bits, rounded to the nearest integer. %
}
\label{fig-matrix}
\end{center}
\end{figure}

\begin{figure}[t]
\begin{center}
\includegraphics{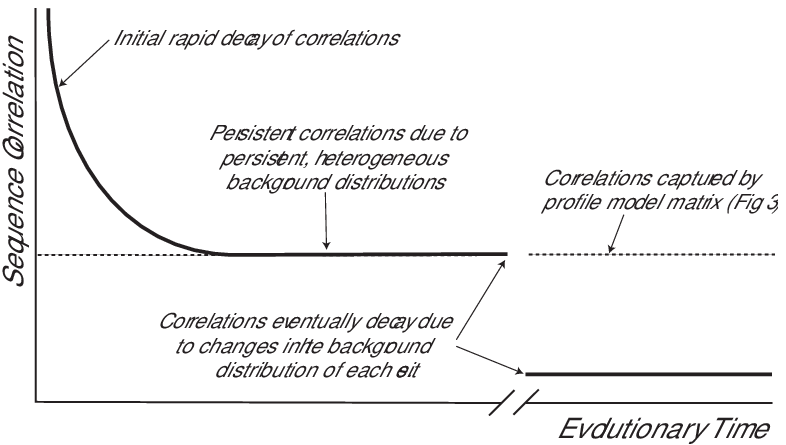}
\caption{
Schematic representation of the decay of sequence correlation with evolutionary time. There is an initial rapid reduction in correlation on the time scale of single residue substitutions.  Under the standard Markov model, this exponential decay would continue. However, under the profile model the correlations instead limit towards a plateau value, due to the heterogeneity of the background amino acid distribution. These are the correlations captured by the  substitution matrix of Fig.~\ref{fig-matrix}. Finally, over a second, much longer time scale the sequence correlations decay towards insignificance  due to changes in the site-specific background distribution. 
}
\label{fig-decay}
\end{center}
\end{figure}

The correlations between pairs of homologous residues can be summarized by an amino acid substitution matrix, $S$, whose entries represent the log  probability of observing the homologous pair of amino acids $q_{ij}$ in a properly aligned pair of homologous proteins, against the probability $p_i p_j$ of independently observing the residues in unrelated sequences \citep{Altschul-1991-JMB}.
\begin{equation}
	S_{ij} = \frac{1}{\lambda}\log \frac{q_{ij}}{p_i p_j}
\label{submatrix}	
\end{equation}
\noindent
Units of one third bits  are traditional for substitution matrices (base 2 logarithm, $\lambda=1/3$, $\approx \frac{1}{10}$ digits), although the scaling is arbitrary.
Assuming conditionally independent replacements, we can directly construct the large time limit substitution matrix from the background probability distribution, $P(\theta)$. 
Fortunately, this large distribution has previously been investigated and parameterized  by fitting many columns from multiple alignments of  homologous protein sequences to a mixture of Dirichlet distributions
\citep{Karplus-1995-TN-Regularizers, DurbinEddy1998}.
Fig.~\ref{dirichlet} displays a projection of the {\tt dist.20comp}  parameterization  \citep{Karplus-1995-TN-Regularizers} and Fig.~\ref{fig-matrix} displays the corresponding substitution matrix.
 The mathematical details of matrix construction are given below.

At shorter evolutionary times there is a significant chance that no mutation has occurred at all, resulting in an enhanced probability of amino acid conservation. Let $c$ be the probability of zero mutation events, then the substitution matrix, adjusted for the possibility of zero mutations, is 
\begin{equation}
S_{ij}(c) = \log \frac{c p_i \delta_{ij} + (1-c) q_{ij} }{ p_i p_j} ,
\label{submatrix2}
\end{equation}
\noindent where $\delta_{ij}$ is the Kronecker delta function. A reasonable default model for the conservation probability $c$ would be to assume that substitutions are Poissonian. Then $c = \exp(-t/\tau)$, where $\tau$ is the mean time between replacements. Note that although replacement is Markovian (albeit zeroth order), the dynamic decay of residue correlations at a position is not,  due to the heterogeneity of the amino acid background at that position (an unobserved hidden variable).

Eq.~\ref{submatrix2}  gives the log odds of aligned residues, given the inter-sequence divergence. Conversely, given a prior on the parameter $c$ and a fixed alignment we can invert Eq.~\ref{submatrix2} and estimate the divergence between sequences.  
Conserved residues indicate small divergences and unconserved pairs argue for large divergences, although different pairs are weighted differently.

\begin{figure*}[t!]
\begin{center}
\includegraphics{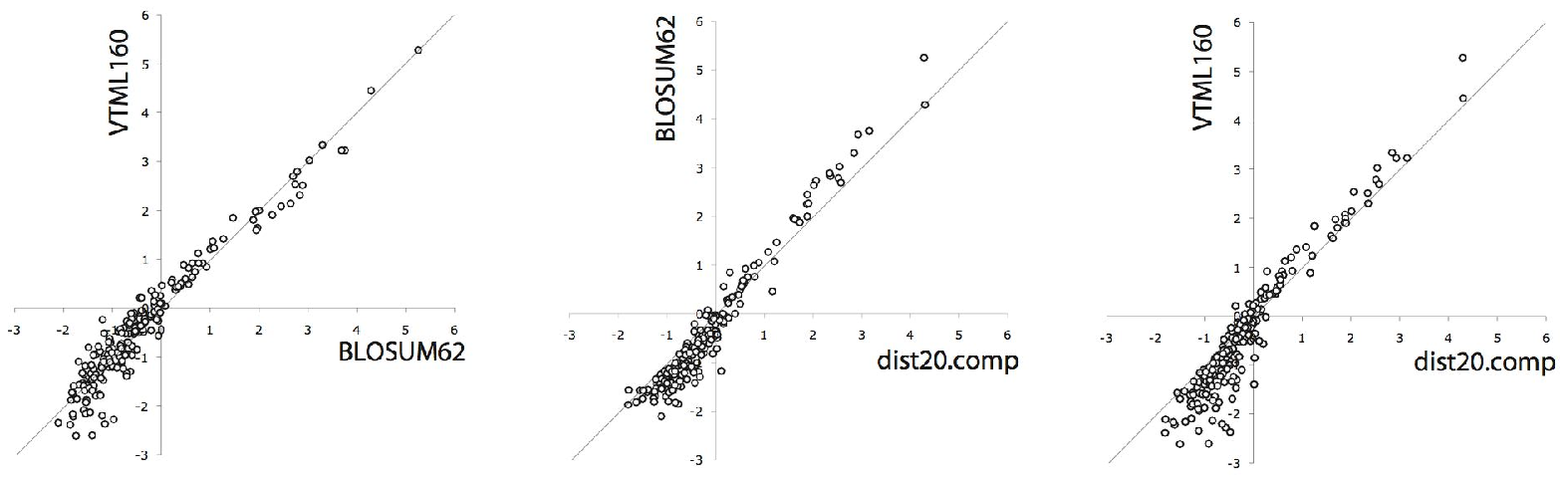}
\caption{Comparison between the bit scores of BLOSUM62 \citep{HenikoffHenikoff-1992-PNAS}, VTML160 \citep{Muller-2002-MolBiolEvol} and {\tt dist.20comp}  (Fig.~\ref{fig-matrix}) substitution matrixes.  All three matrices reflect similar levels and patterns of sequence divergence, but have been derived using very different approaches. The BLOSUM matrices are empirical, the VTML family are based upon the Markov model of amino acid substitution, and the {\tt dist.20comp} matrix is based upon the conditionally independent substitution model.  
}
\label{fig-comparison}
\end{center}
\end{figure*}

As evolution proceeds, the background distribution of a site may itself change, due, for example, to a change in structure of that part of the protein. This will result in a loss of homology signal under our model, and it may no longer be possible to align the diverged residues, nor to recognize them as homologs. This is schematically illustrated in Fig.~\ref{fig-decay}.

Various families of substitution matrices have been developed, included PAM, BLOSUM and VTML. Different members of the same family represent different degrees of sequence divergence.
In principle, we should match the divergence inherent in the substitution matrix to the divergence of the pair of sequences we wish to align \citep{Altschul-1993-JMolEvol}. However, this is computationally expensive, and, in practice, a single matrix is chosen based on its ability to align remote homologs, on the grounds that matching close homologs is relatively easy \citep{Brenner-1998-PNAS}. Under the Markov model, the chosen matrix  has no particular significance. On the other hand, under our model there is a natural, non-trivial, long time limit matrix (Fig.~\ref{fig-matrix}; Eq.~\ref{submatrix2}, $c=0$). This matrix represents the sequence correlations at any time after the first few mutations, and before the underlying amino acid background itself diverges.
(Fig.~\ref{fig-decay})

\begin{figure}[t]
\begin{center}
\includegraphics{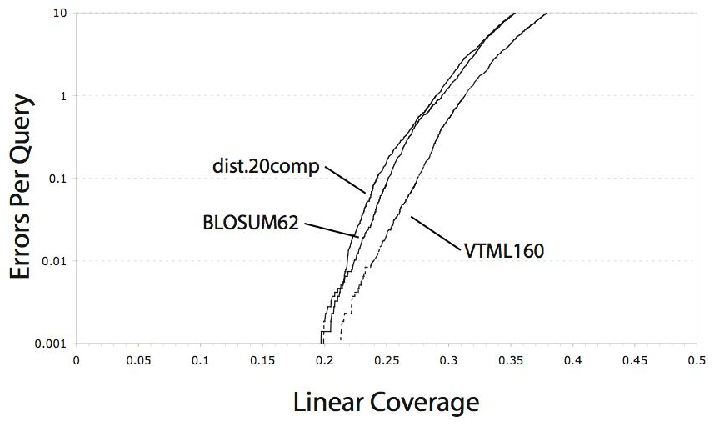}
\caption{The substitution matrices, BLOSUM62, VTML160 and {\tt dist.20comp}, are of  comparable effectiveness under Green and Brenner's (2002) 
evaluation of pairwise remote homology detection. A set of
about 2800 sequences (none of which share more than 40\% sequence identity) are collated from the SCOP (Structural Classification of Proteins) database (version 1.57) \citep{Murzin-1995-JMB, Brenner-2000-NAR-ASTRAL}. SCOP reliably clusters these sequences into groups of homologs  using structural information. Each sequence is matched against the dataset using Smith-Waterman alignment \citep{SmithWaterman1981}, a particular substitution matrix and appropriate gap penalties \citep{Green-2002-IEEE}. The results are shown as a plot of errors per query against (linearly normalized) coverage, the average fraction of true homologs that are found for each sequence. There is a trade off between accuracy and coverage; the bottom right of the above graph is ideal; high coverage with few errors. 
We used Bayesian bootstrap resampling to estimate that the standard deviation of the coverage is about 0.02  at 0.01 errors per query. \citep[Price, Crooks, Green \& Brenner, Unpublished]{Green-2002-IEEE,Zachariah2004}.
Thus, there is a statistically significant, but relatively small and (in practice) unimportant variation in homology coverage between the three matrices. 
Note that both BLOSUM and VTML matrices have been directly trained upon pairwise alignment data, and may therefore be favored in this pairwise alignment test.}
\label{fig-CVE}
\end{center}
\end{figure}

Figures~\ref{fig-comparison} and \ref{fig-CVE} demonstrate that the {\tt dist.20comp}  matrix  represents a similar level of evolutionary divergence, and similar patterns of substitution as  BLOSUM62 and VTML160, two substitution matrices commonly used for pairwise sequence alignment and remote homology detection. These three matrices have been created using very different evolutionary models; the {\tt dist.20comp} matrix is based upon our heterogeneous background/independent substitution model, and the   {\tt dist.20comp}  background distribution is,
in turn, derived from the columns of many multiple alignments of homologous protein sequences; the popular BLOSUM matrices are empirically derived from the BLOCKS database of reliable protein sequence alignments \citep{HenikoffHenikoff-1992-PNAS,Henikoff-2000-NAR-Blocks}; and the classic PAM \citep{Dayhoff1978} and modern VTML \citep{Muller-2002-MolBiolEvol} matrix families are explicitly based upon the Markovian model of amino acid replacement. In a recent evaluation of pairwise remote homology detection, the VTML160 matrix was found to be more effective than any other VTML, PAM or BLOSUM matrix.  \citep{Green-2002-IEEE} However, as can be seen in figure~\ref{fig-CVE}, the difference in remote homology detection ability of the three matrices is relatively small.

In summary, the important sequence correlations can be adequately explained by assuming conditionally independent replacements drawn from  background distributions that vary from site-to-site, but are stable over evolutionary time-scales. The standard, Markovian model of amino acid replacement is unnecessary, overly complicated and inconsistent with observed substitution patterns. 

This alternative, heterogeneous background, independent substitution model may be particularly useful for simultaneous sequence alignment and phylogenetic tree reconstruction,  since it is  necessary to align pairs of close homologs at the leaves, and multiply align many remote homologs at the interior nodes of the tree.  Therefore,  a  simple (yet realistic) evolutionary dynamic that is consistent across a wide range of divergence times, and that leads naturally to sequence-sequence, sequence-profile and profile-profile alignment algorithms, may be advantageous.


\section*{Mathematical Details}

A collection of homologous residues can be represented by a 20 component canonical amino acid count vector, $n= \{n_1, \ldots,n_{20}\}$. The total number of counts can be 1, if the observation is taken from a single sequence, or many if the collection represents an entire column of a multiple sequence alignment or some other related set of residues. 

In general, we wish to estimate whether two collections of homologous residues are 
related, given that detectably homologous residues are drawn from the same background amino acid distribution.
The appropriate test statistic is the log odds of sampling the two amino acid count vectors from the same, but unobserved, background distribution, against the probability of independently sampling the two count vectors from different distributions.
\begin{equation}
S(n^1,n^2) = \log \frac{P(n^1, n^2)}{P(n^1) P(n^2)}
\label{lod}
\end{equation}
\noindent

The probability of independently sampling a particular collection of homologous residues $n$ from the background amino acid profile from which those residues are drawn, $\theta = \{\theta_1, \ldots, \theta_{20}\}$, follows the multinomial distribution;
\begin{eqnarray}
P(n|\theta) &=& {\mathcal M}(n| \theta) \nonumber \\
&=&    \frac{1}{M(n)}  \prod_{i=1}^{20} \theta_{i}^{n_i},
\quad M(n) = \frac{ \prod_i n_i !}{(\sum_i n_i)!} .
\end{eqnarray}
\noindent This is the multivariant generalization of the common binomial distribution.

The probability distribution of background distributions $P(\theta)$ has been studied and measured by collating columns from many multiple protein sequence alignments. Since this is a very large, multi-modal probability it is necessary to parameterize the distribution into a convenient representation. Typically, a mixture of Dirichlet distributions is used 
\citep{Karplus-1995-TN-Regularizers, Sjolander-1996-CABIOS, DurbinEddy1998}. 
\begin{equation}
P(\theta) = \sum_{k=1,m} \rho_k {\mathcal D}(\theta|\alpha^k)
\end{equation}
\noindent The $m$ mixture coefficients $\rho_k$ sum to one.
The $k$th Dirichlet distribution is itself parameterized by the 20 component (canonical amino acids) non-negative vector $\alpha^k$,
\begin{equation}
{\mathcal D}(\theta | \alpha) = \frac{1}{Z(\alpha) }
\prod_{i=1}^{20} \theta_i^{(\alpha_i-1)},
\quad Z(\alpha) =  \frac{\prod_i \Gamma(\alpha_i)}{\Gamma(A)} ,
\end{equation}
\noindent 
where $A=\sum_i \alpha_i$. 

Dirichlet mixtures are used to model the background probability partially because Dirichlet distributions are naturally conjugate to the multinomial distribution (and therefore mathematically convenient) and partially because a Dirichlet mixture can approximate the true distribution with a reasonably small number of parameters. The underlying assumption is that the probability distribution is smooth, but lumpy. (See Fig.~\ref{dirichlet})

If we do not know the particular background from which the observations have been drawn, then we must average over all backgrounds to find the probability of observing a particular count vector; 
\begin{eqnarray}
P(n) 
&=& \int d\theta \; P(n |\theta) P(\theta) ,\nonumber\\
&=& \int d\theta \; {\mathcal M}(n|\theta) \sum_k \rho_k {\mathcal D}(\theta | \alpha^k) ,
\nonumber\\
&=& \int d\theta \sum_k  \rho_k 
 \frac{ \prod_{i=1}^{20} \theta_{i}^{n_i} }{M(n)} 
\frac{\prod_{i=1}^{20} \theta_i^{(\alpha^k_i-1)}}{Z(\alpha^k) }
\nonumber\\
	    &=& \sum_k \rho_k  \frac{1}{Z(\alpha^k)}\frac{1}{M(n)} \int d\theta 
	    \prod_{i=1}^{20} \theta_i^{(n_i +\alpha^k_i-1)}  ,
	    \nonumber\\
	    &=&  \sum_k \rho_k  \frac{ Z(n + \alpha^k) }{Z(\alpha^k)M(n)}.
\label{P1}
\end{eqnarray}
The last line follows because the product in the previous line is an unnormalized Dirichlet with parameters $(n+\alpha^k)$. Therefore, the integral over $\theta$ must be equal to the corresponding Dirichlet normalization constant,  $Z(n + \alpha^k)$.
The final result is a mixture of multivariate negative hypergeometric distributions \citep{JohnsonKotz-1969}.
The negative hypergeometric is  an under-appreciated distribution \citep[e.g. Eq.~11.23,][]{DurbinEddy1998} which bares the same relation to the hypergeometric as the negative binomial does to the binomial distribution. The multivariant generalization appears in this case as the combination of a Dirichlet and a multinomial.
Confusingly, the negative hypergeometric distribution is sometimes called the inverse hypergeometric, an entirely different distribution, and vice versa.

The probability of independently sampling two count vectors, $n^1$ and $n^2$, from the same undetermined background is 
\begin{eqnarray}
P(n^1, n^2) &=&  \int d\theta \; P(n^1 |\theta) P(n^2 |\theta) P(\theta) ,\nonumber\\
&=& \int d\theta \; {\mathcal M}(n^1|\theta) {\mathcal M}(n^2|\theta)
			 \sum_k  \rho_k {\mathcal D}(\theta | \alpha^k) 
\nonumber \\ 
&=& \sum_k \rho_k  \frac{1}{Z(\alpha^k)} \frac{1}{M(n^1)}\frac{1}{M(n^2)} 
\times \nonumber \\
& & \qquad \int d\theta
	    \prod_{i=1}^{20} \theta_i^{(n^1_i+n^2_i +\alpha^k_i-1)} 	    
\nonumber \\
	   &=&  \sum_k \rho_k  \frac{ Z(n^1+n^2 + \alpha^k) }{Z(\alpha^k)M(n^1)M(n^2)}
\label{P2}
\end{eqnarray}

Combing Eqs.~\ref{P1} and~\ref{P2} with the log likelihood ratio, Eq.~\ref{lod}, generates a generic profile-profile sequence alignment score that is valid whether the number of counts is small or large. 
\begin{equation}
S(n^1,n^2) =
\log \frac{
 \sum\limits_k \rho_k  \frac{ Z(n^1+n^2 + \alpha^k) }{Z(\alpha^k)M(n^1)M(n^2)}
}{ \sum\limits_k \rho_k \frac{    Z(n^1 + \alpha^k) }{Z(\alpha^k)M(n^1)}  \,
 \sum\limits_k  \rho_k \frac{    Z(n^2 + \alpha^k) }{Z(\alpha^k)M(n^2)}
}
\label{final}
\end{equation}

 For the particular case that one of the count vectors contains only a single observation this score reduces to the standard sequence-profile score frequently used by hidden Markov model protein sequence alignment \citep{Sjolander-1996-CABIOS}. This is inevitable, since the underlying mathematics is the same.

If both count vectors contain only a single observation, then this profile-profile score reduces to a pairwise substitution matrix. Note, given that $n^1_x = \delta_{xi}$ and $n^2_x = \delta_{xj}$ (where $\delta_{xj}$ is a Kronecker delta function), then all but the $j$th element of the product $Z(\delta_{xj}+\alpha^k)/Z(\alpha^k) $ cancels.
Thus,%
  \begin{eqnarray}
  S_{ij} &=& \log \frac{ q_{ij} }{ p_{i} p_{j} } 
\nonumber  \\  
 p_i &=& \sum_k \rho_k \frac{\alpha^k_i}{A^k}
 \nonumber
 \\
  q_{ij} &=& \left\{
 \begin{array}{rl} 
 \displaystyle
\sum\limits_k \rho_k \frac{\alpha_i^k \alpha_j^k}{ A^k ( A^k +1)}
 & i\neq j
 \\
\displaystyle
\sum\limits_k \rho_k \frac{\alpha_i^k (\alpha_i^k+1)}{ A^k ( A^k +1)}
& i = j
 \end{array} \right.
 \label{pairwise} 
 \end{eqnarray}

Applying Eq.~\ref{pairwise} to  the 20 component Dirichlet mixture {\tt dist.20comp}  generates the pairwise substitution matrix illustrated in Fig.~\ref{fig-matrix}.

An interesting feature of this model is that it provides a unified homology match score for  sequence-sequence,  sequence-profile and profile-profile alignment (Eq.~\ref{final}).  As far as we are aware, this profile-profile score has not been evaluated in a profile-profile alignment algorithm, although it is a natural generalization of the established hidden Markov model profile-sequence score. However, in the large sample limit Eq.~\ref{final} reduces to the Jensen-Shannon  divergence between the two empirical amino acid distributions, a measure that has shown some promise in profile-profile alignment \citep{Yona2002,Edgar2004,Marti-Renom2004}.

\subsection*{Acknowledgments}
We would like to thank  Richard E. Green,  Marcus Zachariah, Robert C. Edgar and Emma Hill for helpful discussions 
and suggestions. This work was supported by the National Institutes of Health 
(1-K22-HG00056). GEC received funding from the Sloan/DOE postdoctoral fellowship in computational molecular biology. SEB is a Searle Scholar (1-L-110).


 \end{document}